\documentclass[sigconf, nonacm]{acmart}

\newcommand\vldbdoi{XX.XX/XXX.XX}

\newcommand\vldbavailabilityurl{URL_TO_YOUR_ARTIFACTS}

\usepackage{caption}
\usepackage{subcaption}
\usepackage{circledsteps}
\usepackage{algorithm}
\usepackage[noend]{algorithmic}

\usepackage[framemethod=default]{mdframed}
\mdfdefinestyle{NR}{skipabove=\topskip,skipbelow=\topskip,align=center,innerleftmargin=.25cm,linecolor=black,linewidth=1pt,backgroundcolor=lightgray!30}
\newmdenv[style=NR]{NR}

\newcommand{\cy}{\pgfkeys{/csteps/inner color=black}\pgfkeys{/csteps/outer color=yellow}\pgfkeys{/csteps/fill color=yellow}}
\definecolor{ao(english)}{rgb}{0.0, 0.5, 0.0}
\newcommand{\cg}{\pgfkeys{/csteps/inner color=white}\pgfkeys{/csteps/outer color=ao(english)}\pgfkeys{/csteps/fill color=ao(english)}}
\newcommand{\cb}{\pgfkeys{/csteps/inner color=white}\pgfkeys{/csteps/outer color=blue}\pgfkeys{/csteps/fill color=blue}}
\newcommand{\cred}{\pgfkeys{/csteps/inner color=white}\pgfkeys{/csteps/outer color=red}\pgfkeys{/csteps/fill color=red}}

\usepackage{xcolor}

\newcommand{\where}{\textbf{\textit{where}}}
\newcommand{\why}{\textbf{\textit{why}}}
\newcommand{\how}{\textbf{\textit{how}}}

\begin{document}

\title[ChaTEAU]{ChaTEAU: A Universal Toolkit for Applying the \textsc{Chase}}

\author{Tanja Auge}
\email{tanja.auge@uni-rostock.de}
\affiliation{%
\institution{University of Rostock}
\city{Rostock}
\country{Germany}
}

\author{Nic Scharlau}
\author{Andreas Görres}
\author{Jakob Zimmer}
\affiliation{%
\institution{University of Rostock}
\city{Rostock}
\country{Germany}
}

\author{Andreas Heuer}
\email{andreas.heuer@uni-rostock.de}
\affiliation{%
\institution{University of Rostock}
\city{Rostock}
\country{Germany}
}

\begin{abstract}
What do applications like semantic optimization, data exchange and integration, answering queries under dependencies, query reformulation with constraints, and data cleaning have in common? 
All these applications can be processed by the \textsc{Chase}, a family of algorithms for reasoning with constraints. While the theory of the \textsc{Chase} is well understood, existing implementations are confined to specific use cases and application scenarios, making it difficult to reuse them in other settings. ChaTEAU overcomes this limitation: It takes the logical core of the \textsc{Chase}, generalizes it, and provides a software library for different \textsc{Chase} applications in a single toolkit.


\end{abstract}

\maketitle

%

\pagestyle{empty}

\section{Introduction}
\label{sec:intro}
The \textsc{Chase} is a widely applicable technique for reasoning with constraints. It takes a parameter $\ast$ and an object $\bigcirc$ as input, and forms a result that corresponds to the combination of both. In this way, the parameter is incorporated into the object, so that $\ast$ is explicitly contained in $\bigcirc$, denoted $\text{\textsc{Chase}}_\ast(\bigcirc) = \text{\textcircled{$\ast$}}$. The versatile applicability of the \textsc{Chase} is due to the fact that one can pass different types of objects and parameters as input. Instead of considering queries and instances separately (as other implementations do), ChaTEAU generalizes both to a \textsc{Chase} object. Similarly, the \textsc{Chase} parameter in ChaTEAU generalizes dependencies, queries, and views by treating them uniformly as logic formulas.

Applying the \textsc{Chase} to instances \cite{BMKPTMS17} and queries \cite{DPT99} behaves in a similar manner, because the structure of queries and instances is also quite similar. However, existing \textsc{Chase} tools such as PDQ~\cite{BLT14}, Llunatic \cite{GMPS20}, or Graal \cite{BLMRS15} are limited to specific use cases, e.g., semantic optimization, data cleaning and exchange, or query answering with existential rules. 
These different use cases can be reduced to the processing of instances and queries.
With ChaTEAU (\textsc{\textbf{Cha}se} for \textbf{T}ransforming, \textbf{E}volving, and \textbf{A}dapting databases and queries, \textbf{U}niversal Approach) we have developed and implemented a universal \textsc{Chase} tool that abstracts instances and queries to a general \textsc{Chase} object and parameter. The software, examples, and further information are available at our Git repository\footnote{Git repository: \url{https://git.informatik.uni-rostock.de/ta093/ChaTEAU-demo}}.

The uniform treatment of \textsc{Chase} use cases and variants in ChaTEAU makes it ideal for embedding it in different applications, e.g., for data exchange, data cleaning, or query reformulations with constraints.
For specific applications, additional extensions such as provenance or a second \textsc{Backchase}-phase may be necessary. These extensions are being added gradually to ChaTEAU and can be selected individually depending on the target use case. We are currently integrating \textbf{\textit{where}}-, \textbf{\textit{why}}- and \textbf{\textit{how}}-provenance \cite{HDL17} to provide provenance-supported applications as well.

\paragraph{\textbf{Structure of the article:}} 
Section ~\ref{sec:generalization} describes our \textsc{Chase} generalization; Section \ref{sec:features} discusses the ChaTEAU implementation. Finally, the ChaTEAU demonstration and GUI are presented in Section \ref{sec:demo}, using a concrete example. 

\section{Generalization of the \textsc{Chase}}
\label{sec:generalization}
Recall that the \textsc{Chase} modifies a given object $\bigcirc$, called \textit{\textsc{Chase} object}, by incorporating a parameter $\ast$ (the \textit{\textsc{Chase} parameter}), which we can write as $\text{\textsc{Chase}}_\ast(\bigcirc) = \text{\textcircled{$\ast$}}$. While $\bigcirc$ can represent both queries and instances, we understand $\ast$ as set of constraints formalized as (s-t) tgds and/or egds. An \textit{equality generating dependency (egd)} is a formula of the form $\forall \textbf{x}(\phi(\textbf{x}) \rightarrow (x_1 = x_2))$. A formula of the form $\forall \textbf{x}(\phi(\textbf{x}) \rightarrow \exists \textbf{y} : \psi(\textbf{x}, \textbf{y}))$ is called \textit{(source-to-target) tuple generating dependency ((s-t) tgd)} with $\phi$ (\textit{body}) and $\psi$ (\textit{head}) conjunctions of atomic formulas over a source and target schema, respectively. If the source and target schemas are the same, the constraint is simply a tgd. As their names suggest, egds and tgds derive new equalities and new tuples (with $\exists$-quantified variables), respectively.

\begin{figure}[t]
\centering
\includegraphics[width=0.4\textwidth]{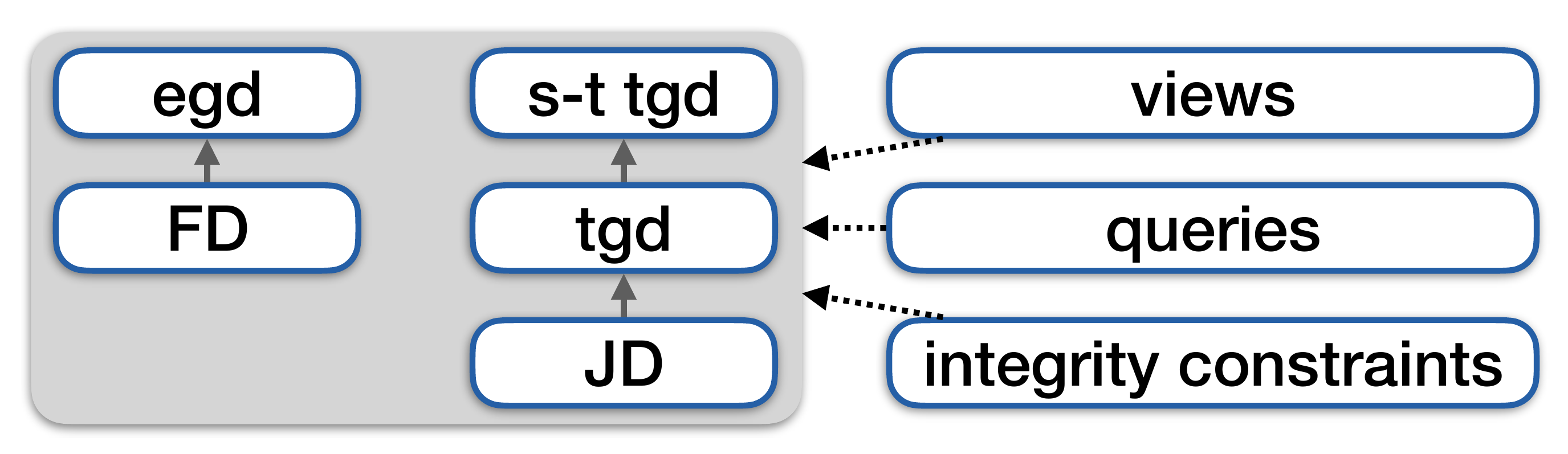}
\caption{Constraints hierarchy and how they are formalized}
\label{fig:formalisations}
\end{figure}

First approaches to extend the \textit{Standard \textsc{Chase}} \cite{BMKPTMS17} to arbitrary objects and parameters can be found in \cite{AH19}. Note that the \textsc{Chase} parameter $\ast$ either represents intra-database dependencies (as tgds or egds) or inter-database dependencies (as s-t tgds). The hierarchy in Figure \ref{fig:formalisations} shows how other dependencies can be represented as either (s-t) tgds or egds. 

The \textsc{Chase} object $\bigcirc$ is either a query $Q$ or a database instance $I$. In both cases, variables and null values can be replaced by other variables and null values or constants. The variable substitution rules depend on certain conditions. Let's have a closer look at them.

\paragraph{\textbf{\textsc{Chase} Parameter}}
The \textsc{Chase} parameter $\ast$ consists of a set of dependencies $\Sigma$ in the form of egds or (s-t) tgds. These are generalizations of the classical \textit{functional dependencies (FD)} and \textit{join dependencies (JD)}. Any condition that can be written as a set of (s-t) tgds and egds can be used as a \textsc{Chase} parameter. This includes views, 
queries, and integrity constraints as seen in Figure \ref{fig:formalisations}.

\paragraph{\textbf{\textsc{Chase} Object}}
The database tuple $\textsc{student}(3,\text{'Max'},\text{'Math'})$ and the query atom  $\textsc{student}(y_\text{id},x_\text{name},\text{'Math'})$ are very similar in structure. A \textsc{Chase} object is an abstraction of both. The tuple consists of constants ($c_i$) and null values ($\eta_1$) while the expression contains (implicitly) $\forall$-quantified variables ($x_i$), $\exists$-quantified variables ($y_i$) and constants ($c_i$).

A database \textit{instance} $I$ over schema $R$ consists of finite relations  $R^I_1, ..., R^I_k$, where each relation $R_i^I$ has the same arity as the relation symbol $R_i$. Each tuple $(x_1, ..., x_n)$ in $R^I_i$ consists of constants $c_i$ or null values $\eta_i$. 
A \textit{conjunctive query} is a first-order formula of the form $\exists y : \phi(x,y) \rightarrow \psi(x)$ with $\phi(x,y)$ a conjunction of logic atoms (the \textit{body}) and $\psi(x)$ a single atom (the \textit{head}). The terms in $\phi$ are $\forall$-quantified or $\exists$-quantified variables, or constants. The head $\psi$ must not contain $\exists$-variables.

A query $Q$ can be transformed into a \textit{frozen instance} $I_Q$, in which each atom of $Q$'s body is represented as a tuple in $I_Q$~\cite{DNR08}. There are different ways to deal with the variables in $Q$. Often $\exists$-variables are transformed into null values and $\forall$-variables are treated as \textit{labeled null} values or special constants. For the transformation of a conjunctive query $Q$ into a generalized instance, the atoms in the body have to be written as generalized tuples. We create tuples with $\exists$- and $\forall$-variables, e.g., as follows: 
\begin{eqnarray*}
Q & = & \exists y_\text{id}: \textsc{student}(y_\text{id},x_\text{name},\text{'Math'}) \rightarrow (x_\text{name}) \\
I_Q & = & \{\textsc{student}(y_\text{id},x_\text{name},\text{'Math'})\}.
\end{eqnarray*}
These generalized tuples can be extended to a generalized instance, as described in Definition \ref{def:instance}. While ChaTEAU can handle general s-t tgds, here we focus on queries, which can be seen as s-t tgds with a single atom in the head.

\begin{definition}
\label{def:instance}
Let $I$ be an instance and $Q$ a conjunctive query. A \textit{generalized instance} is either:
\begin{itemize}
\item a set of (conventional) relations $R^I_1, ..., R^I_k$, i.e., where tuples consist of constants and null values, or
\item a set of generalized tuples, consisting of the atoms of $\phi(x,y)$,  with constants, $\forall$-variables, and $\exists$-variables. 
\end{itemize}
\end{definition}

\paragraph{\textbf{The \textsc{Chase} for Generalized Instances}}
Due to the different kinds of \textsc{Chase} objects $O_i$, the \textsc{Chase} steps $I_{O_i} \rightarrow I_{O_{i+1}}$ have to be generalized too, when using (s-t) tgds and egds. In ChaTEAU, we thus extend the Standard \textsc{Chase} \cite{BMKPTMS17} to a \textit{\textsc{Chase} for generalized instances} (see Algorithm \ref{alg:CHASE}). 

The main task of the \textsc{Chase} is to infer new facts. To this end, we need to find mappings (\textit{homomorphisms}) between the dependencies $\Sigma$ and the \textsc{Chase} object $O$. Using these, the \textsc{Chase} maps a set of dependencies $\Sigma$ into $O$. The result is a modified \textsc{Chase} object $O'$. Between $O$ and $O'$ we also have a homomorphism. 

\begin{definition}
Let $\phi(\textbf{x})$ be the body and $\psi(\textbf{x},\textbf{y})$ the head of a dependency. Let $I_{O_1}$ and $I_{O_2}$ be generalized instances. We define the following possible substitution rules:
\begin{enumerate}
\item a constant can be mapped to itself: $c_i \mapsto c_i$
\item a null value can be mapped to a constant, itself, or another null value: $\eta_i \mapsto c_j \mid \eta_i \mid \eta_j$
\item an $\exists$-variable can be mapped to a constant, itself, a null value, or another $\exists$- or $\forall$-variable: \\
(a) $y_i \mapsto c_j \mid \eta_j \mid y_i \mid x_j$ \hspace{1cm} (b) $y_i \mapsto c_j \mid y_i \mid y_j \mid x_j$
\item a $\forall$-variable can be mapped to a constant, a null value, itself, or another $\forall$- or $\exists$-variable: \\
(a) $x_i \mapsto c_j \mid \eta_j \mid x_i \mid y_j$ \hspace{1.05cm} (b) $x_i \mapsto c_j \mid x_i$
\end{enumerate}
Now a \textit{homomorphism} $h: \phi(\textbf{x}) \rightarrow I_{O_1}$ is a mapping that satisfies (1) and (4a); a \textit{homomorphism} $h: \psi(\textbf{x},\textbf{y}) \rightarrow I_{O_2}$ is a mapping that satisfies (1), (3a), and (4a); and a \textit{homomorphism} $h: I_{O_1} \rightarrow I_{O_2}$ is a mapping that satisfies (1), (2), (3b), and (4b).
\end{definition}

Tgds insert new tuples to instances or add atoms to a query body. These changes are created by applying a particular homomorphism, called \textit{trigger}, from the tgd-body to the generalized instance. Egds equate variables by applying a homomorphism from the egd-body to the generalized instance. An \textit{active trigger} is one that (1) creates new tuples or expressions by applying an (s-t) tgd, or (2) leads to a new equation of variables or null values by applying an egd. 

\begin{definition}
\label{def:trigger}
A \textit{trigger} is a homomorphism $h$ from a dependency body to a generalized instance, i.e., $h : \phi(\textbf{x}) \rightarrow I_O$. 
An \textit{active trigger} is a trigger that satisfies for a
\begin{itemize}
\item[(1)] tgd: no extension of $h$ to an homomorphism $\psi(\textbf{x},\textbf{y}) \rightarrow I_O$,
\item[(2)] egd: $h(x_1) \neq h(x_2)$.
\end{itemize}
\end{definition}
If no new tuples or equals are created during a \textsc{Chase} execution, the trigger is not active. However, if an $\exists$-variable is contained in the tgd-head, we always have an active trigger.
These variables will map to new null values or $\exists$-variables, depending on the type of \textsc{Chase} object. 

We extend the Standard \textsc{Chase} to generalized instances. This new \textsc{Chase} version modifies a generalized instance by a sequence of \textsc{Chase} steps until all dependencies are satisfied.

\begin{definition}
Let $h : \phi(\textbf{x}) \rightarrow I_{O_i}$ be an active trigger for a dependency $\sigma$ and a generalized instance $I_{O_i}$. The modification of $I_{O_i}$ to $I_{O_{i+1}}$ by applying $\sigma$ under $h$ is called \textit{\textsc{Chase} step}. 
\end{definition}

\begin{definition}
Let $\Sigma$ be a set of dependencies and $I_{O_0}$ a generalized instance. The (finite) \textit{\textsc{Chase} for generalized instances} is a finite sequence of \textsc{Chase} steps $I_{O_i} \rightarrow I_{O_{i+1}}$ ($0 \le i \le n$) with 
\begin{itemize}
\item $I_{O_n} = \bot$ (\textsc{Chase} fails),
\item $I_{O_n} = I_{O_{n+1}}$, i.e. exists an homomorphism $h: I_{O_n} \rightarrow I_{O_{n+1}}$ with $h(z_j) = z_j$ and $z_j \in \{c_j, \eta_j, x_j, y_j\}$.
\end{itemize}
\end{definition}

Finally, the result calculated with the \textsc{Chase} on generalized instances must be interpreted. Applying an egd, the \textsc{Chase} on instances fails if different constants are matched to each other, and returns $\bot$, whereas the \textsc{Chase} on queries returns $\emptyset$. The \textsc{Chase} result on $Q$ corresponds to the transformation of $I_{Q_n}$ into a new query $Q'$. For this, the tuples of $I_{Q_n}$ form a conjunction of atoms in the body of $Q'$. The query head is formed by applying the composition of all homomorphism collected during the \textsc{Chase}-execution. Thus, the \textsc{Chase} implemented in ChaTEAU works on arbitrary s-t tgds.

\begin{algorithm}
\caption{\textsc{Chase} for generalized instances ($\Sigma$, $I_{O_0}$)}
\label{alg:CHASE}
\begin{algorithmic}[1] 
\REQUIRE set of dependencies $\Sigma$, a database instance $I_{O_0}$
\ENSURE modified database instance $I_{O_n}$
\WHILE{$I_{O_j} \neq \bot$ \AND $I_{O_{j-1}}\neq I_{O_j}$}
\FORALL{trigger $h$ for $\sigma \in \Sigma$}
\IF {\textcolor{red}{$h$ is an active trigger}}
\IF {$\sigma$ is a tgd}
\STATE extending $h$ and adding new tuples to instance $I_{O_j}$ 
\ELSIF {$\sigma$ is an egd} 
\IF {values compared are different constants}
\STATE $I_{O_{j+1}} = \bot$
\ELSE 
\STATE substitute null values and variables by other null values, variables, or constants 
\ENDIF
\ENDIF
\ENDIF
\ENDFOR
\ENDWHILE
\end{algorithmic}
\end{algorithm}

\section{ChaTEAU}
\label{sec:features}
ChaTEAU runs on different types of \textsc{Chase} parameters and objects, which are automatically recognized and processed accordingly. Different constraint and termination checks are applied. The results of these tests and of the individual \textsc{Chase} steps are stored in a log.

\paragraph{\textbf{Input and Output}}
Input and output of instances, queries, and constraints to ChaTEAU is done through special XML files. The input file defines the schema, consisting of the relation schemas and dependencies (the \textsc{Chase} parameter), as well as an instance or a query (the \textsc{Chase} object). 

\paragraph{\textbf{Termination}}
Inserting tuples that contain null values may cause non-termination of execution. This happens whenever tgds interact and trigger each other and generate new null values each time they are used. Conditions that guarantee a fixed point of a \textsc{Chase} sequence exists are called \textit{termination conditions}, and several of these can be found in \cite{GST11}. ChaTEAU  implements five of them: rich acyclicity, weak acyclicity, safety, acyclicity, and acyclicity with egd rewriting.


Acyclicity is a very powerful condition based on constraint rewriting \cite{GST11}, which is extensible and easy to implement. Commonly found is the test for weak acyclicity, such as in Llunatic \cite{GMPS20} or Graal \cite{BLMRS15}. It can be expanded to rich acyclicity and safety without much effort. Additionally, we decided to implement acyclicity with egd rewriting to better handle the problem of egds which are ignored in most termination criteria \cite{GST11}.


\begin{figure}[b]
\centering
\includegraphics[width=.95\columnwidth]{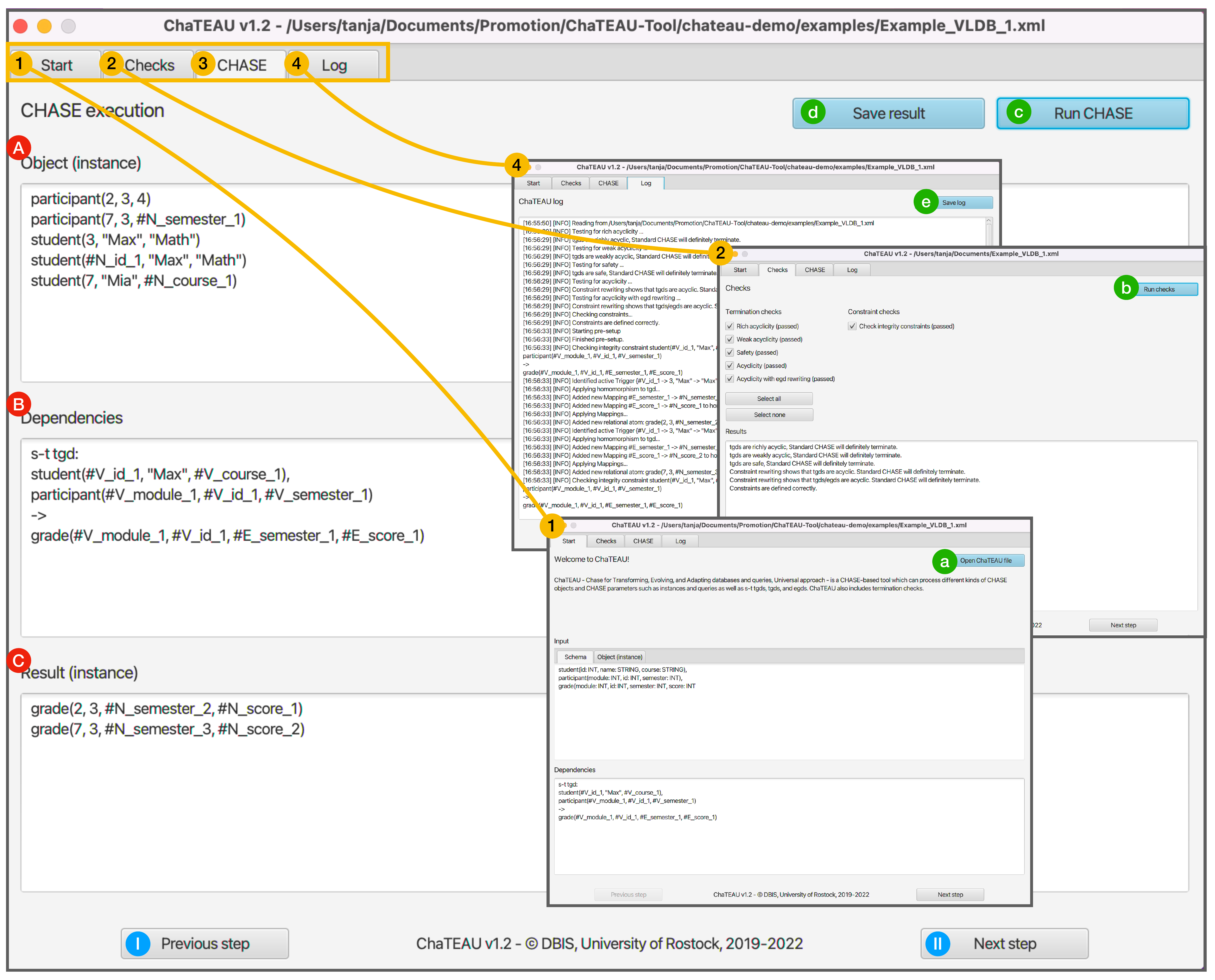}
\caption{Overview of ChaTEAU (zoomable in pdf version)}
\label{fig:ChaTEAUoverview}
\end{figure}

\paragraph{\textbf{API}}
ChaTEAU is a stand-alone application for the \textsc{Chase}. It is implemented as a Maven project and can easily be accessed through the GUI presented in Section \ref{sec:demo}. In addition, ChaTEAU can be accessed via its API, making it easy to employ it as building block or library for developing other \textsc{Chase}-based applications. 

In our research project \textit{ProSA} \cite{AH19}, e.g., we combine provenance management with the \textsc{Chase} algorithm to compute the inverses of evaluation queries. 
ProSA is a tool that employs a variant of the \textsc{Chase} on instances, called \textsc{Chase}\&\textsc{Backchase}, so ChaTEAU is called twice: once for the \textsc{Chase} and once for the \textsc{Backchase}~\cite{Aug20}.

\paragraph{\textbf{Further development of ChaTEAU}}
We extended ChaTEAU further by adding additional features. For example, \where-, \why-, and \how-provenance \cite{HDL17} --- as needed for ProSA --- have already been implemented. However, provenance is not part of the \textsc{Chase} itself, but is required for integration into ProSA or other applications. Therefore, this feature can only be used via the API, and not through the GUI. Other extensions such as a general attribute-constant comparisons or the integration of negated atoms and relations are still in progress. We are also continuing work on new ChaTEAU applications, as presented in \cite{AH19}.

\section{Demonstration}
\label{sec:demo}


The ChaTEAU GUI (see Figure \ref{fig:ChaTEAUoverview}) is divided into four tabs: \texttt{Start} (\cy\Circled{\textbf{1}}), \texttt{Tests} (\Circled{\textbf{2}}), \texttt{\textsc{Chase}} (\Circled{\textbf{3}}) and \texttt{Log} (\Circled{\textbf{4}}). The demo starts by opening a special XML file called \textit{ChaTEAU file} (\cg\Circled{\textbf{a}}). It contains the \textsc{Chase} parameters as a set of dependencies and the \textsc{Chase} object. The ChaTEAU system automatically determines whether the object is a query or an instance and adjusts the associated tags (\cred\Circled{\textbf{A}} and \Circled{\textbf{B}}) accordingly. Before executing the \textsc{Chase} in the third step (\cy\Circled{\textbf{3}}), termination and constraint checks are performed (\Circled{\textbf{2}}). A variety of tests can be selected and  executed sequentially. The relevant log can be found on the last tab (\Circled{\textbf{4}}).

We will navigate in ChaTEAU using the two buttons \texttt{Previous step} (\cb\Circled{\textbf{I}}) and \texttt{Next \!\!step} (\Circled{\textbf{II}}) at the bottom of the window. Within the upper right corner of each tab, there is an option to save or execute something (\cg\Circled{\textbf{a}} -- \Circled{\textbf{e}}).



\subsection{\textsc{Chase} on Instances}
Even though the \textsc{Chase} on instances and queries works the same in ChaTEAU, we present them as separate use cases. 
We start with an example for chasing instances. Both examples can be found in the corresponding demo repository.

\paragraph{\textbf{Start}} 
Instead of manually entering formulas, the fields for the \textsc{Chase} object (\texttt{Input}) and parameter (\texttt{Dependencies}) are automatically generated from the selected XML file. 

We consider an instance (object) and a query (parameter) formalized as an s-t tgd that generates a table of grades from a student and a participant table. Thus, known attributes like \textsc{id} and \textsc{module} are adopted and new null values for \textsc{semester} and \textsc{score} are introduced. In addition, all students not named Max are filtered out.

\begin{NR}
\footnotesize
\textbf{Instance:}
\vspace{-0.38cm}
\begin{eqnarray*}
\hspace{3.125cm} \scriptstyle \{\text{participant}(2,3,4), \ \text{ participant}(7,3,\#N\_\text{semester}\_1), \\
\scriptstyle \text{student}(3, \text{'Max'}, \text{'Math'}), \ \text{ student}(\#N\_\text{id}\_1, \text{'Max'}, \text{'Math'}), \ \text{student}(7, \text{'Mia'}, \#N\_\text{course}\_1) \} 
\end{eqnarray*}
\footnotesize \textbf{Dependencies:}
\begin{eqnarray*}
\scriptstyle \text{participant}(\#V\_\text{module}\_1, \#V\_\text{id}\_1, \#V\_\text{semester}\_1), \
\text{student}(\#V\_\text{id}\_1, \text{'Max'}, \#V\_\text{course}\_1) \\
\scriptstyle -> \text{grade}(\#V\_\text{module}\_1, \#V\_\text{id}\_1, \#E\_\text{semester}\_1, \#E\_\text{score}\_1)
\end{eqnarray*}
\end{NR}

\paragraph{\textbf{Termination}} 
ChaTEAU implements five common termination tests, from which the user can choose. When the button \texttt{Run select\-ed checks} (\Circled{\textbf{b}}) is pressed, all checks will be run. The process can be repeated as often as desired (e.g. with different termination checks). 

In our example, the \textsc{Chase} terminates according to all five criteria. Also the constraint check is successful.

\begin{NR}
\scriptsize
tgds are richly acyclic -> Standard \textsc{Chase} will definitely terminate. \\
tgds are weakly acyclic -> Standard \textsc{Chase} will definitely terminate. \\
tgds are safe -> Standard \textsc{Chase} will definitely terminate. \\
Constraint rewriting shows that tgds are acyclic -> \textsc{Chase} will definitely terminate. \\
Constraint rewriting shows that tgds/egds are acyclic -> \textsc{Chase} will definitely terminate. \\
Constraints are defined correctly.
\end{NR}

\paragraph{\textbf{\textsc{Chase} Execution}} 
The key part of ChaTEAU is the \textsc{Chase} application in the third tab. In addition to the input (\cred\Circled{\textbf{A}} and \Circled{\textbf{B}}), we also see the \textsc{Chase} result (\Circled{\textbf{C}}) here, which can be saved (\cg\Circled{\textbf{d}}). Despite negative termination tests, the \textsc{Chase} can still be executed. In this case, an alert appears. If the \textsc{Chase} is still running, it can be stopped by clicking the {Start \textsc{Chase}} button (\cg\Circled{\textbf{c}}) again, which is now labeled as a {Stop} button. The \textsc{Chase} steps can be reviewed in the log (\cy\Circled{\textbf{4}}). 

Our result instance by applying \textsc{Chase} matches the result of the SQL query \texttt{SELECT * FROM participant NATURAL JOIN student WHERE name = 'Max'} to the instance defined above:
\begin{NR}
\footnotesize
\textbf{Result (instance):}
\begin{eqnarray*}
& & \scriptstyle \text{grade}(7,3,\#N\_\text{semester}\_2,\#N\_\text{score}\_1), \ \text{grade}(2,3,\#N\_\text{semester}\_3,\#N\_\text{score}\_2)
\end{eqnarray*}
\end{NR}

\paragraph{\textbf{Logging}} The \textsc{Chase} execution is finished after three steps (\cy\Circled{\textbf{1}}~--~\Circled{\textbf{3}}). The log offers additional information such as the results of the single \textsc{Chase} steps after application of a (s-t) tgd or egd and details regarding the termination checks carried out. The log is especially suitable for debugging and is saved using \cg\Circled{\textbf{e}}.

\subsection{\textsc{Chase} on Queries}
Thus ChaTEAU provides, depending on the CHASE object $\bigcirc$ an instance or query extended by the specified parameter $\ast$. We continue with an example for chasing queries.

\paragraph{\textbf{Start}}
We consider a query (object) and a constraint (parameter) formalized as egd. The egd replaces the $\exists$-quantified variable $\#E\_\text{course}\_1$ with the $\forall$-quantified variable $\#V\_\text{course}\_1$ by equating the attributes $\#V\_\texttt{course}\_1$ and $\#V\_\text{course}\_2$. 

\begin{NR}
\footnotesize
\textbf{Query:}
\begin{eqnarray*}
\scriptstyle \text{student}(\#V\_\text{id}\_1, \#V\_\text{name}\_1, \#E\_\text{course}\_1), \ \scriptstyle \text{student}(\#E\_\text{id}\_1, \#V\_\text{name}\_1, \#V\_\text{course}\_1) \\
\scriptstyle -> \ (\#V\_\text{id}\_1, \#V\_\text{name}\_1, \#V\_\text{course}\_1)
\end{eqnarray*}
\footnotesize
\textbf{Dependencies:}
\begin{eqnarray*}
\scriptstyle \text{student}(\#V\_\text{id}\_1, \#V\_\text{name}\_1, \#V\_\text{course}\_1), 
\ \text{student}(\#V\_\text{id}\_1, \#V\_\text{name}\_1, \#V\_\text{course}\_2) \\
\scriptstyle -> \ \#V\_\text{course}\_1 \ = \ \#V\_\text{course}\_2 
\end{eqnarray*}
\end{NR}

\paragraph{\textbf{Termination and Logging}}
Both termination and logging behave as for instances. All six checks are satisfied.

\paragraph{\textbf{\textsc{Chase} Execution}}
Applying the CHASE provides the substitution of $\#E\_\text{course}\_1$ described above. As a result, the body has changed, while the head remains the same.

\begin{NR}
\footnotesize
\textbf{Result (query):}
\begin{eqnarray*}
\scriptstyle \text{student}(\#V\_\text{id}\_1, \#V\_\text{name}\_1, \#V\_\text{course}\_1), \ \text{student}(\#V\_\text{id}\_2, \#V\_\text{name}\_1, \#V\_\text{course}\_2)) \\
\scriptstyle \quad -> \ (\#V\_\text{id}\_1, \#V\_\text{name}\_1, \#V\_\text{course}\_1)
\end{eqnarray*}
\end{NR}


\section{Conclusion}
\label{sec:conclusion}
The \textsc{Chase} implemented in ChaTEAU can be applied to queries and instances. Cha\-TEAU combines these two approaches by incorporating a set of views, queries, and dependencies formalized as (s-t) tdgs and egds, called \textsc{Chase} parameter, into a general \textsc{Chase} object. For this, queries can be interpreted as frozen instances. This also means that a different treatment of queries and instances is no longer needed. ChaTEAU thus offers a versatile implementation of a family of reasoning algorithms, which can be easily integrated into other \textsc{Chase}-based applications such as ProSA \cite{AH19}. 

\begin{acks}
We thank all the students who were involved in developing Cha\-TEAU. In particular, we would like to mention Martin Jurklies, Fabian Renn, Florian Rose, Michael Albus, Eduard Buch, Lukas Görtz, Moritz Hanzig, Eric Maier, Rocco Flach and Chris Röhrs. We also thank Bertram Ludäscher for comments and suggestions.
\end{acks}



\end{document}